\documentclass[a4paper,fleqn]{cas-dc}

\usepackage[numbers]{natbib}

\def\tsc#1{\csdef{#1}{\textsc{\lowercase{#1}}\xspace}}
\tsc{WGM}
\tsc{QE}
\tsc{EP}
\tsc{PMS}
\tsc{BEC}
\tsc{DE}

\newcommand*{\email}[1]{\href{mailto:#1}{\nolinkurl{#1}} } 

\begin{document}
\let\WriteBookmarks\relax
\def\floatpagepagefraction{1}
\def\textpagefraction{.001}
\shorttitle{GNOME and LBM Model for Ocean Oil Spills}
\shortauthors{Zhang et~al.}

\title [mode = title]
{GNOME and LBM Model Evaluation on Ocean Oil Spill
Far-Field Impacts to Highly Sensitive Areas
}                      



\author[1]{Zhanyang Zhang}[type=author,
                        auid=001,bioid=1] 
\address[1]{Doctoral Program in Computer Science, The Graduate Center, City University of New York, 365 5th Avenue, New York, New York 10016, USA}

\author[2] {Tobias Sch{\"a}fer}[type=author,
                        auid=002,bioid=2]
\address[2]{Doctoral Program in Physics, The Graduate Center, City University of New York, 365 5th Avenue, New York, New York 10016, USA}

\author[1]{Michael E. Kress}[type=author,
                        auid=003,bioid=3]


\begin{abstract}
In case of an ocean oil spill, there are certain areas, e.g. shrimp farms, which are highly sensitive to small amounts of oil pollution while they are geographically far from the spill.  We investigate the Lattice Boltzmann Method (LBM) and GNOME, a tool developed and used by NOAA, in terms of far-field impacts to sensitive areas. We present our simulation results of both models in limited scale (a sub area of Gulf of Mexico) under the same oil spill condition using real ocean current data from the Unified Wave Interface-Coupled Model (UWIN-CM).  Our study shows that the kinetic theory based LBM model outperforms the stochastic particle based GNOME model in accuracy and computation time due to their fundamental difference in representation of oil pollution advection and diffusion mechanisms. We propose LBM as a viable alternative to the Lagrangian particle calculation component of GNOME in modeling the far-field impacts to highly sensitive areas. 
\end{abstract}

\begin{graphicalabstract}
\scalebox{0.75}{\includegraphics{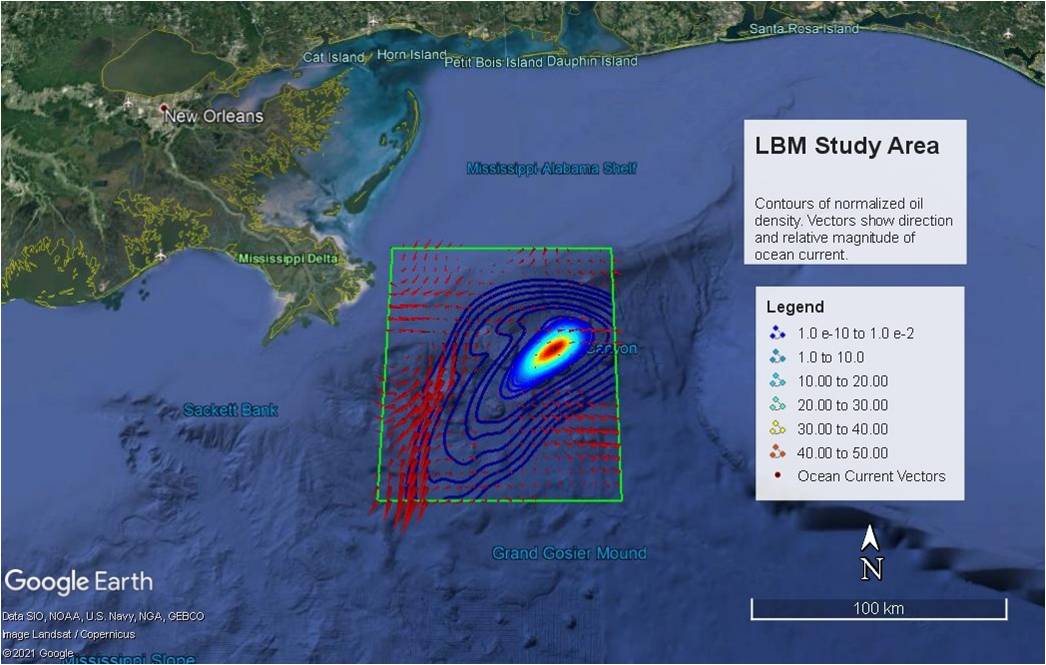}}
\end{graphicalabstract}

\begin{highlights}
\item Ocean oil spills have devastating impacts on coastal environment and fishery industry.
\item Shrimp farms are highly sensitive to small amounts of oil pollution in term of far-field impacts.
\item GNOME is based on a model using stochastic Lagrangian particles to track oil pollutants.
\item The Lattice Boltzmann method (LBM) is a kinetic model using a density distribution to track oil pollutants. The LBM is computationally efficient, accurate, and well-suited to assess the far-field impact of oil spills.
\item The LBM can be a viable alternative to the Lagrangian particle calculation component of GNOME,
in particular when modeling the far-field impacts to highly sensitive areas.
\end{highlights}


\maketitle

\section{Introduction}
Ocean oil spills have devastating impacts on marine eco\-systems and human society in the surrounding coastal areas. The 2010 Deep Water Horizon spill in the Gulf of Mexico lasted 87 days and was estimated to have released over 3 million barrels of oil. It impacted over 1,600 miles of coastline, killed over 8,000 marine animals/seabirds and caused direct economic loss from fishing and tour industries estimated at tens of billions of dollars \cite{graham:bp2011}. 

There are certain areas that are highly sensitive to small amount of oil pollutants, such as shrimp farms.  The toxic effect of hydrocarbons can lead to mortality, if the levels exceed the threshold concentration despite the fact that these areas can be geographically far from the center of oil spills \cite{bodkin_esler_rice_matkin_ballachey:2014, Gracia:2020}. Understanding and being able to predict these far-field impacts of ocean oil spill is critical for EPA first response teams and shrimp farmers to allocate resource to prevent loss and mitigate risk \cite{IPIECA:2004}.  

How much oil (measured as oil concentration level) and how long it takes for surface oil to propagate to the highly sensitive areas are challenging problems. The transport of oil spilled into the ocean is a complex process that depends in a critical way on the current, wind, temperature and chemical composition of the oil and seawater \cite{mishra:oilspillwethering}.  We focus our study on investigating the feasibility and effectiveness of using the Lattice Boltzmann Method for solving the Advection-Dif\-fusion Equation (LBM-ADE) to model and simulate ocean oil spill far-field impacts to address the aforementioned challenge problems. 

We developed a prototype model and simulation in limited scale, a sub area of the Gulf of Mexico, with assimilation of real data from the UWIN-CM ocean current model \cite{uwin-cm:ocean-model}. We conducted two simulation experiments in comparison of the LBM-ADE model vs GNOME (General NOAA Operational Modeling Environment), a modeling tool which the Office of Response and Restoration’s (ORR) Emergency Response Division has been using to predict the possible oil transports on an ocean surface \cite{noaa:gnome}. GNOME is an ADE-based tool with Lagrangian particles, which relies on the accuracy of the ocean surface current velocity data to produce quality results. Our study shows the kinetic theory based LBM model outperforms the stochastic particle based GNOME model in accuracy and computation time due to their fundamental difference in representation of oil pollution advection and diffusion mechanisms. We propose the LBM as a viable alternative to the Lagrangian particle calculation component of GNOME in modeling the far-field impacts to highly sensitive areas. 

There are many publications regarding using LBM to model and solve ocean flow problems. Wolf-Gladrow’s work \cite{gladrow:lbm} used the LBM to solve the linearized Munk Problem \cite{munk:oceanmodell}. In another LBM application of ocean models, Nuraiman  \cite{nuraiman:lbm-ocean} used a 1D Shallow Water Equation representation of the Navier Stokes Equations coupled to the 2D Navier Stokes Equations to form an LBM model using the Bhatnagar, Gross and Krook (BGK) kinetic theory \cite{cercignani:bgk}. But to our knowledge there have been only a few studies of oil spill tracking using the LBM. One of the most comprehensive studies was done by \cite{maslo:lbmoill} and showed good agreement between simulated results and satellite observations from an oil spill in the Gulf of Beirut on July 15, 2006. Their LBM model used a two relaxation parameter technique to facilitate numerical stability. In addition, a flux limiter computational technique was used to resolve sharp numerical boundaries, which led to negative densities. Further, an interpolation technique was used to permit a non-square lattice to resolve the flow along the elongated coastline studied. In addition, Ha and Ku \cite{haku:lbmoil} used an LBM model to simulate an advective-diffusion formulation of the spread of an oil slick on the sea surface and confirmed the functionality of their model. Further, Li et al. \cite{lmk:lbmcde} solve the 2D convection-diffusion equation using the LBM. Other advection-diffusion equation solutions are presented by Dedits et al. \cite{dpsv:lbm-ade}. While not specifically studying oil transport, Li and Huang \cite{lihuang:lbm-shallow-water} used a coupled LBM formulation of the Shallow Water Equation and Contamination Concentration Transport. Excellent agreement was obtained between numerical predictions and analytical solutions in the pure diffusion problem and convection–diffusion problem. Banda and Seaid \cite{banda:lbm-shallow-water} also developed an LBM  model to solve shallow water equations as the depth-averaged incompressible Navier-Stokes equations with conservation of momentum under the assumption that the vertical scale is much smaller than any typical horizontal scale and the pressure is hydrostatic. Then they apply their shallow water model to simulate pollutant transport in the Strait of Gibraltar.

Our literature review shows most of ocean oil spill and contamination transport models are based on the ADE. In addition to the above cited research, GNOME is an ADE-based tool, which has been used to predict the possible route, or trajectory, a pollutant might follow on the surface of water.  It relies on the accuracy of the ocean surface current velocity field to produce quality results. Also, GNOME is based on simulating an ensemble of stochastic particles such that the accuracy depends crucially on the number of particles in the simulation. In contrast, LBM computes the distribution function directly and is expected to be computationally more efficient in situations where GNOME would require an exceedingly large number of particles in order to provide an accurate solution.

During the first phase of our research, we developed an LBM-ADE model and simulation that is capable of providing numerical solutions as an LBM-ADE solver. To validate the model, we performed a benchmark test using a Gaussian Hill concentration with a simplified velocity field first. However, the ocean surface current constitutes a much more complex velocity field that is temporal-spatial dependent. We tested the LBM-ADE solver against a Finite Differential Method (FDM) ADE solver using a perturbation of the Taylor-Green velocity field. To the best of our knowledge, no such benchmark has been done in the past for an LBM-ADE model using a velocity field as complex as the perturbed Taylor-Green field. Our first phase study shows the LBM-ADE model achieves great results in comparisons vs analytical solution in Gaussian Hill case and vs the FDM-ADE solution in Taylor-Green case  \cite{zzhang:ams2020}.

Afterward, we conducted a benchmark study of LBM-ADE and GNOME against an analytical ADE solution and a comparison study of the LBM-ADE model vs the GNOME model in prediction the surface transport of spilled oil  in the same sub area of the Gulf of Mexico, with assimilation of real data from the UWIN-CM ocean current model. 
We achieved accurate results using Gaussian Hill concentrations with two ocean current scenarios: linear ocean current and real ocean current from (UWIN-CM). These results suggest that the LBM-ADE is a promising model that is capable of predicting spilled oil transport on an ocean surface \cite{zzhang:wsc2020}.

\section{Methodology}

\subsection{GNOME Model}
GNOME \cite{noaa:gnome} is an ADE-based model using Lagrangian particles, referred as Lagrangian elements (LEs), to represent oil pollutants. All the LEs trajectories collectively represent the path of oil transport on the ocean surface.  In a simplified ocean oil spill scenario without considering weathering process, each LE is moved independently by two forces, namely advection and diffusion. Furthermore GNOME assumes the advection and diffusion processes are independent of each other.  
GNOME models the advection process using a forward Euler scheme. Assume at time step $t$, a LE is at the point $p(x,y,t)$. It calculates the LE position after one time step $t+\Delta t$ at the point $p(x+\Delta x, y+\Delta y, t+\Delta t)$ using  the velocity field for each element.

Diffusion is modeled as stochastic processes where a set of LEs engage a 2-D random walk with a displacement probability such that the mean value remains zero, but the variance grows linearly with time. It has been shown that a long series of random steps will converge to a Gaussian distribution with variance growing linearly with time \cite{csanady:diffusion}. For a given input of diffusion coefficient, GNOME uses a particle random walk simulation, a stochastic process, to approximate the diffusion process.

\subsection{LBM-ADE Model} \label{sec:lbm-ade-model}

\begin{figure}[htb]
{
\centering
\includegraphics[width=0.45\textwidth]{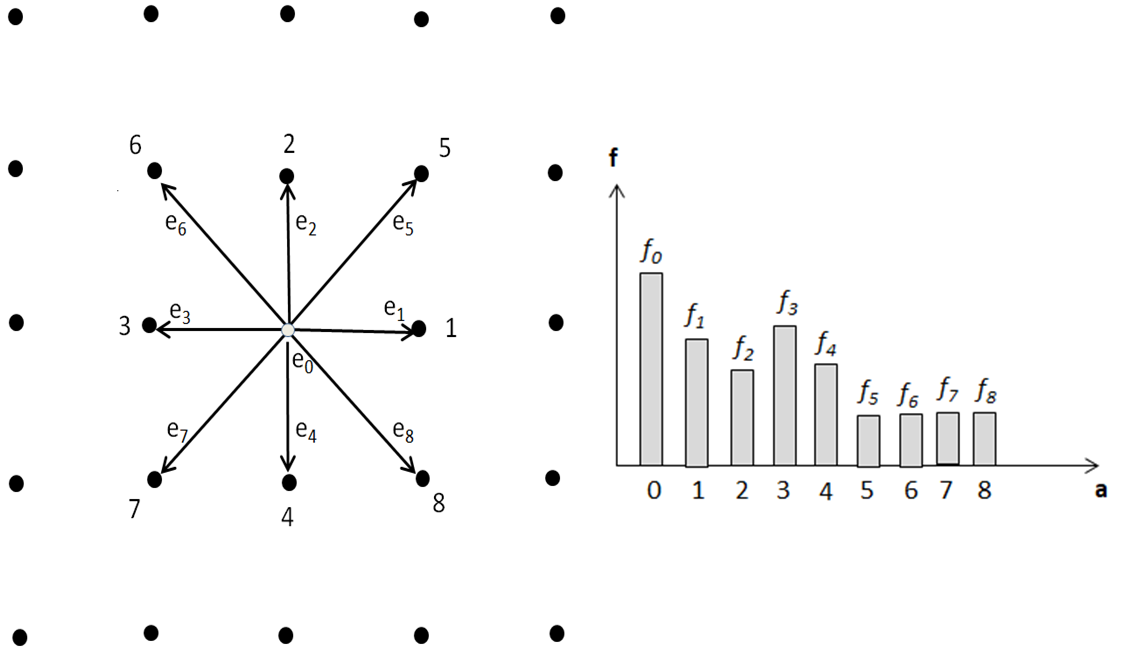}
\caption{The particle distribution function represents the percentage of particles in the velocity bins.\label{fig:lbm-particles}}
}
\end{figure}

We use the LBM to model ocean oil pollutants as a set of particles with certain density and mass located on a virtual grid (lattice) that maps over an area of ocean with boundary conditions representing coastal lines or islands.  This model makes it possible to track particle spatial positions and microscopic momenta from a continuum to just a handful and similarly discrete in distinct steps. Particle positions are confined to the nodes of the lattice. Variations in momentum that could have been due to a continuum of velocity directions and magnitudes and varying particle mass are reduced (in a simple 2D model) to 9 directions and a single particle mass \cite{sukop:lbm-geoscientist}.  Figure \ref{fig:lbm-particles} shows the Cartesian lattice and the velocities $e_a$ (where   a = 0, 1 … 8) is a direction index and $e_0 = 0$ denotes particles at rest. This model is known as D2Q9 as it is 2 dimensional and contains 9 velocities. It can be generalized to a 3 dimensional model as D3Q27 if we replace the lattice in D2Q9 with a cube with length, width and height are one lattice unit.

The next step is to incorporate the single-species distribution function f, which has only nine discrete ‘bins’ instead of being a continuous function. The distribution function can conveniently be thought of as a histogram representing a frequency of occurrence.  For example the shaded area in Figure \ref{fig:lbm-particles} shows a likely oil pollutant propagation pattern after one time step. 
Accordingly, the macroscopic fluid density is defined as:
\begin{equation}    \label{eq:lbm-mass}
\rho = \sum_{a=0}^{8} f_a 
\end{equation}
The macroscopic velocity $u$ is an average of the microscopic velocities $e_a$ 
weighted by the directional densities $f_a$ as defined as:
\begin{equation}    \label{eq:lbm-velocity}
u=\frac{1}{\rho}\sum_{a=0}^{8} f_a e_a 
\end{equation}
This simple equation allows us to pass from the discrete microscopic velocities that comprise the LBM back to a continuum of macroscopic velocities representing the fluid’s motion. 

When incorporating external forces, such as wind, gravity and others, that interact with the ocean water, Equation \eqref{eq:lbm-force}
can be modified as:
\begin{equation}    \label{eq:lbm-force}
u=\frac{1}{\rho}\sum_{a=0}^{8} f_a e_a + \frac{F\Delta t}{2\rho}
\end{equation}
where the first term is the velocity due to mass density redistribution with conservation of momentum and the second term is due to external forces \cite{kkkssv:lbm}.   

Equation \eqref{eq:lbm-force} is a generalization of the LBM that is applicable to both NSE and ADE models. In the LBM-NSE model, Equation 3 provides a mathematical base for developing a velocity projection schema that integrates the ocean surface velocity into a LBM model as an external input, and then uses it to update local equilibrium distribution functions $f^{eq}$.  On the other hand, in the LBM-ADE model, we ignore the first term in \eqref{eq:lbm-force} and only consider the second term as an advective velocity resulting from external forces since ADE only conserves mass, not momentum.

The next steps are streaming and collision of the particles via the distribution function. The simplest approach to approximate the collision can be defined as: 
\begin{equation}    \label{eq:lbm-collison}
f_a(x+e_a\Delta t, t+\Delta t)=f_a(x,t)- \frac{f_a(x,t)-f_a^{eq}(x,t)}{\tau}
\end{equation}
where $\tau$ is a relaxation time used in the BGK  operator. Although they can be combined into a single statement as above, collision and streaming steps must be separated if solid boundaries are present because the bounce back boundary condition is a separate collision. Collision of the fluid particles is considered as a relaxation towards a local equilibrium. \hfill The parameter $\tau$ is a relaxation time to reach equilibrium. \\*
A D2Q9 equilibrium distribution function $f^{eq}$ is defined as:
\begin{equation}    \label{eq:lbm-feq}
f_a^{eq}(x,t)=w_a\rho(x,t)\left(1+3\frac{e_au}{c^2}+\frac{2(e_au)^2}{9c^4}+\frac{3u^2}{2c^2}\right)
\end{equation}
where the weights $w_a=\left(\frac{4}{9},\frac{1}{9},\frac{1}{9},\frac{1}{9},\frac{1}{9},\frac{1}{36},\frac{1}{36},\frac{1}{36},\frac{1}{36}\right)$ and c is the velocity on the lattice, one lattice unit per time step $(lu/\Delta t)$ in the simplest implementation. Note that if the macroscopic velocity u = 0, the equilibrium $f_a^{eq}$ are simply the weights times the fluid density.  

To implement the LBM model as a simulation program, \cite{bao:lbm-fluid} presented an algorithm outline that can be summarized as follows:
\begin{enumerate}
	\item Initialize $\rho$,  $\mu$, $f_a$ and $f_a^{eq}$;
	
	\item Streaming step: move $f_a$ $\rightarrow$  $f_a^*$ in the direction of $e_a$, where $f_a^*$  holds intermediate values of density distribution after the streaming step.
	
	\item Compute macroscopic $\rho$ and $\mu$ from $f_a^*$ using above equations \eqref{eq:lbm-mass} and
	\eqref{eq:lbm-velocity};
	
	\item Compute $f_a^{eq}$ using
	equation \eqref{eq:lbm-force};
	
	\item Collision step: calculate the updated distribution function  using  equation \eqref{eq:lbm-collison}:  
    $f_a=f_a^*-\frac{f_a^*- f_a^{eq}}{\tau}$;
	
	\item Repeat steps 2 to 5.

\end{enumerate}

During the streaming and collision step, the boundary nodes require some special treatments for the distribution functions in order to satisfy the imposed macroscopic boundary conditions.  
The LBM as described has been shown to be second order accurate in time and space to the 2D incompressible Navier Stokes Equations by \cite{kkkssv:lbm} and separately by \cite{gladrow:lbm}.   

LBM is a kinetic theory based modeling technique that can provide numerical solutions for a range of flow problems whose underline physics are governed  by NSE and/or ADE. When the NSE and the ADE are applied in near incompressible fluids, they can be expressed as equations  \eqref{eq:lbm-nse-pde} and \eqref{eq:lbm-ade-pde}. The Navier Stokes Equations are given by
\begin{equation}    \label{eq:lbm-nse-pde}
\frac{\partial u}{\partial t}+u\nabla u=-\frac{\nabla p}{\rho}+\nu\nabla ^2 u+F
\end{equation}
where $u$ is fluid velocity, $P$ is fluid pressure, $\rho$ is fluid density, $\nu$  is fluid kinematic viscosity, and $F$ is an external force. The Advection-Diffusion Equation is written as 
\begin{equation}    \label{eq:lbm-ade-pde}
    \frac{\partial C}{\partial t}+u\nabla C=D\nabla ^2 C+q\,
\end{equation}
where $C$ is mass concentration, $D$ is diffusion coefficient (assume isotropic diffusion),  $u$ is fluid velocity as an advection force; and $q$ is a source term.

The LBM-NSE model is defined by the set of equations \eqref{eq:lbm-mass} to \eqref{eq:lbm-feq}.  While the LBM-NSE model conserves both mass and momentum, the LBM-ADE model only conserves mass (referred as concentration $C$).  A LBM-ADE model is defined by the set of equations \eqref{eq:lbm-ade-concentration} to \eqref{eq:lbm-ade-geq} as below:
\begin{equation}    \label{eq:lbm-ade-concentration}
C = \sum_{a=0}^{8} g_a 
\end{equation}
\begin{equation}    \label{eq:lbm-ade-collison}
g_a(x+e_a\Delta t, t+\Delta t)=g_a(x,t)- \frac{g_a(x,t)-g_a^{eq}(x,t)}{\tau_g}
\end{equation}
\begin{equation}    \label{eq:lbm-ade-geq}
g_a ^{eq}(x,t)=w_aC(x,t)\left(1+3\frac{e_a u}{c^2}+\frac{2(e_a u)^2}{9c^4}+\frac{3u^2}{2c^2}\right)
\end{equation}
\begin{equation}    \label{eq:lbm-ade-D}
D_l=c^2 \left( \tau_g - \frac{\Delta t}{2} \right) 
\end{equation}
where the $g_a$ are the directional densities of concentration; $g_a^{eq}$ is the equilibrium density function; $\tau_g$ is the relaxation time and $u$ is a velocity vector due to advection forces, while $e_a$ and $w_a$ are the same as previously defined in \eqref{eq:lbm-velocity} and \eqref{eq:lbm-feq}.

Since the equations for the NSE \eqref{eq:lbm-collison} and equation for the ADE \eqref{eq:lbm-ade-collison} are the same, the algorithm outlined by \cite{bao:lbm-fluid} is also applicable to the ADE. 

\section{Model Configurations and Simulations}

LBM and GNOME are both ADE based modeling tools that share the same governing physics laws that  can be defined as a partial differential equation \eqref{eq:lbm-ade-pde} where  we set $q=0$ except in the initial condition for a one-time  oil spill release. 

While the LBM model takes a kinetic approach and the GNOME model takes a stochastic approach, they both provide numerical solutions that approximate and converge to the analytical solution of equation \eqref{eq:lbm-ade-pde}.
But there are  differences between them which present several challenges in the evaluation of their far-field impacts. To conduct a scientfically sound comparison between LBM and GNOME, we  need to carefully design and configure our experiments taking into consideration the following issues: (i) LBM works with a Eulerian specification of flow field, while GNOME works with a Lagrangian specification of flow field; (ii) LBM uses units of measurement of space and time in lattice units, while simulations in GNOME uses units of measurement in meters and seconds; (iii) LBM uses a square lattice to represent a ocean surface area of $1^{\circ}$ by $1^{\circ}$ in longitude and latitude, while GNOME uses a non-square area of 97.904 km by 111.194 km due to the curvature of the earth's surface in longitude and latitude; and (iv) we need to specify the oil spill volumes and diffusion coefficients that are equivalent in both units of measurement in LBM and GNOME.

\subsection{Far-Field Effect Benchmark}

To assist in analyzing the low density effectiveness of GNOME and LBM-ADE we provide a benchmark example which calculates the elapsed time,$ T_{thresh}$, and location of the maximum Y extent,  $Y_{thresh}$, of the minimum non-zero normalized  oil concentration  from a hypothetical point source spill with a constant linear current,  $ (U_x,U_y)= (0.0,-1.0)$.  For this particular case of a constant advection field, we can calculate the theoretical probability density values 
based on the analytical solution of the corresponding ADE and relate these results to the Monte Carlo analysis of the LEs in GNOME. Details of this calculation are provided in the Appendix. In the case for the far-field effect, the stochasticity parameter $\sigma$ in the underlying probability density is given by $\sigma=1.414$. The non-dimensional scaling parameters are: length $(l=100.0)$,  time $(t = 6.42)$, and velocity,  $v=(0.0,-1.0)$.

GNOME is then run with this configuration and diffusion, $D_g= 297761.0 cm^2/sec$.  The number of LEs at $T_{thresh}$ and $Y_{thresh}$ is simulated and the probability is calculated.  To align the initial LE distribution with the initial Gaussian used in the LBM, GNOME is run for 32 time steps with a zero advective velocity field corresponding to setting a current=(0.0,0.0) .

Next, LBM-ADE is run with equivalent parameters and $\tau_g=0.760$ and  the normalized concentration, $C_l$, at  $T_{thresh}$ and $Y_{thresh}$ is calculated. 

The resulting values of the probability and normalized concentrations are presented in Table~\ref{tab:benchmark-gnome-lbm}, where $T_{thresh}$ includes the additional time steps for the development of the initial  Gaussian distribution.  The results show good agreement among GNOME and the LBM-ADE.

\begin{figure}[htb]
{
\centering
\includegraphics[width=0.45\textwidth]{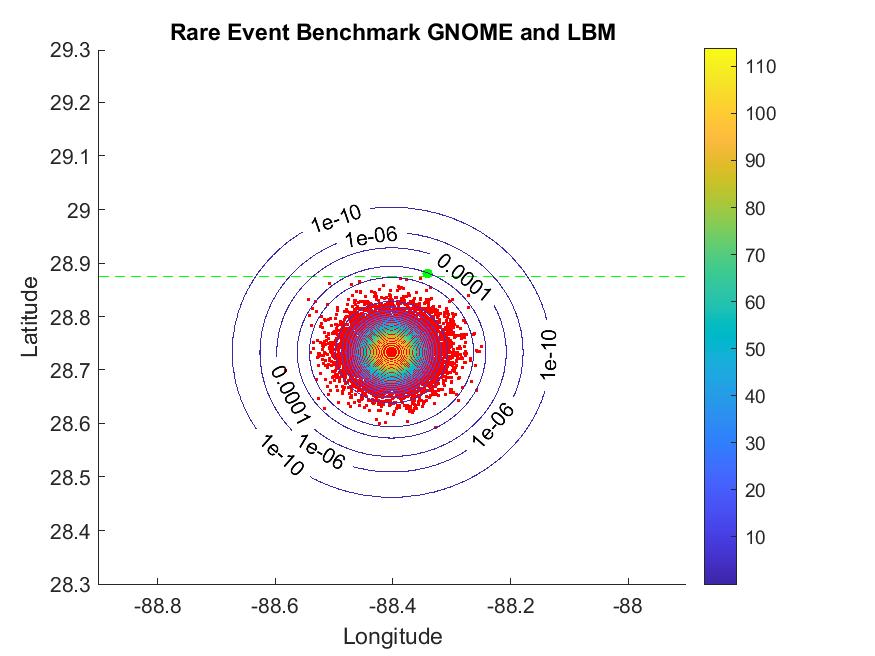}
\caption{The Far-Field Effect Benchmark for both GNOME and LBM.\label{fig:far_field_benchmark}}
}
\end{figure}

Figure \ref{fig:far_field_benchmark} shows the results of the far-field effect, low density, benchmark at $T_{thresh}$ in our computational domain with constant southward velocity at 0.026778 m/s.  The red dots are the locations of the 10,000 LEs  The green dot represents the only LE north of $Y_{thresh}$ shown as the green dashed line.
The contours show the normalized concentration results from the LBM-ADE simulation.  


These results show that LBM-ADE model can provide information of far-field effect with lower oil density to the area beyond the $Y_{thresh}$ while GNOME model can not, since there are no particle presents beyond   $Y_{thresh}$.

\begin{table*}[htb]
\centering
\caption{Benchmark Results Compares Results from GNOME and  LBM-ADE\label{tab:benchmark-gnome-lbm}}
\begin{tabular}{llllll}
\hline
Method & Probability & Total LE Count & $Y_{thresh}$ & $T_{thresh}$ & LEs beyond Threshold\\ \hline

GNOME & $1.00x10^{-4}$ & 10,000 & $28.87^{\circ}$  & 306,000s & 1\\
LBM & $1.20x10^{-4}$ & NA & $1.14x10^{2}$ in lattice length & 296 in lattice time & NA\\
Analytic ADE & $1.22x10^{-4}$ & NA & 7.42 & 6.42 & NA \\
\hline
\end{tabular}
\end{table*}

\subsection{GNOME and LBM ADE Model Configurations and Results Analysis}

In this study, we compare the LBM-ADE solution with GNOME in the same configuration with a velocity field that comes from a real ocean data model (UWIN-CM) to evaluate their effectiveness of far-field impact of ocean oil spill.    

UWIN-CM is a fully coupled atmosphere–wave–ocean system data model.  We used a subset of data from the UWIN-CM ocean model to cover a 1 degree square area of the Gulf of Mexico centered at -88.4 longitude and 28.8 latitude over three days from Feb. 07, 2016 at 16:00:00 until Feb. 10, 2016 at 18:00:00. 
We used bi-linear interpolation spatially to generate a velocity field for the LBM computational domain at each time step while $\Delta t=15$ minutes. The ocean surface velocity field is assimilated in the LBM-ADE model as an advection  velocity at each time step. In GNOME the ocean current data is used in the same way except the velocity data and time steps are stored in a netCDF file, then loaded into the model during model configuration.

We introduce a one-time oil spill as a Gaussian hill in the center of the grid at time step $t=0$ as an initial oil spill in both the GNOME and the LBM-ADE. The volume of a Gaussian hill is a function of $C_0$ and $\sigma_0$ which are calculated to be the volume of oil specified in GNOME.  GNOME uses a point source of 10,000 LEs to represent the oil volume.  Then we let the models run 296 time steps, with $\Delta t=15$ minutes, a total of 74 hours. 

Since in GNOME the initial oil spill is modeled as a point source release, all the LEs are located at the center point of the grid instead of a Gaussian hill distribution as in LBM-ADE. It is interesting to point out that, regardless this initial oil distribution difference, the GNOME concentration diffusion converges to LBM-ADE  after a transient period of approximately 40 time steps.

\begin{figure*}[htb]
{  
\centering
    \includegraphics[width=.45\textwidth]{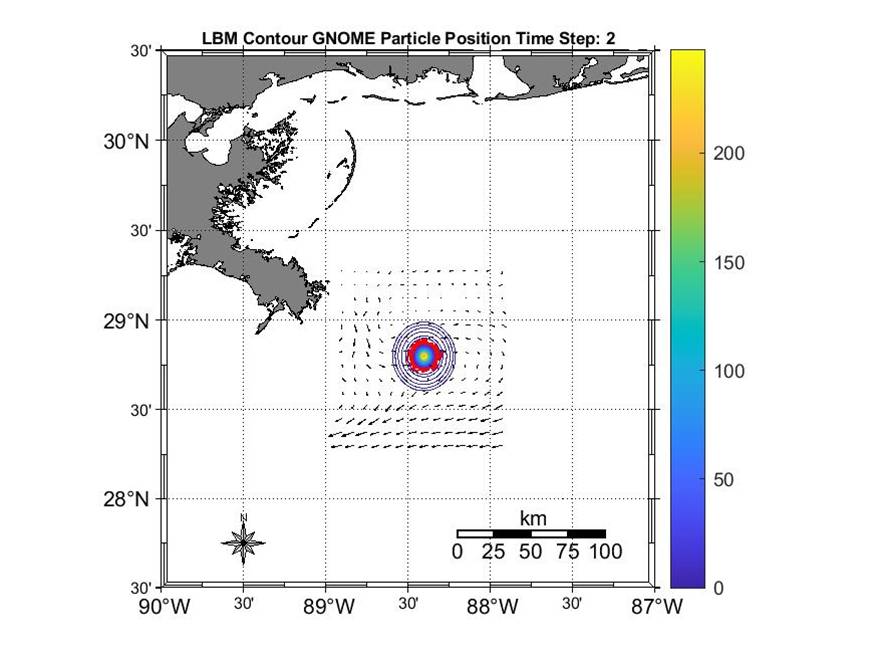} \hfill
    \includegraphics[width=.45\textwidth]{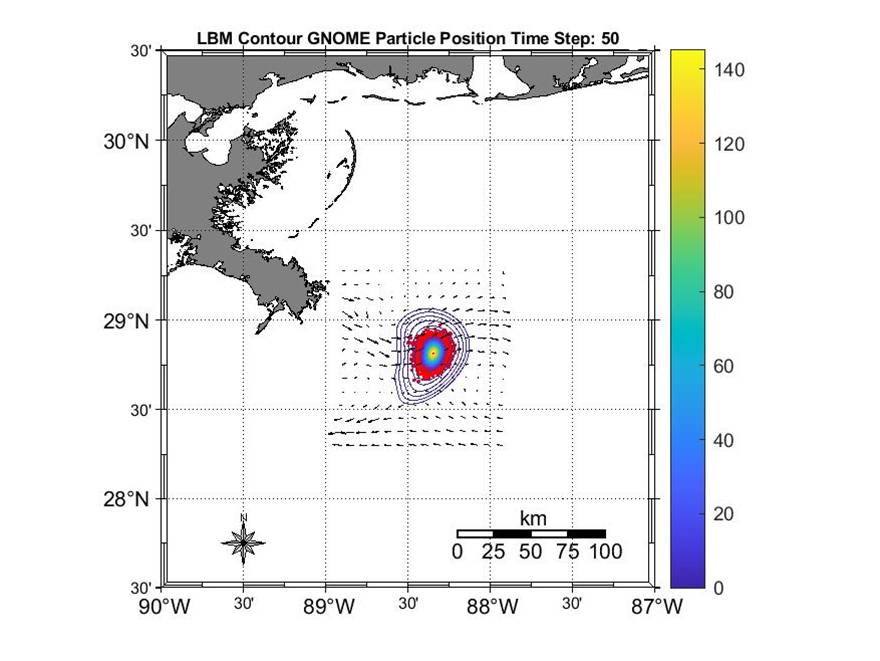} \hfill
    \includegraphics[width=.45\textwidth]{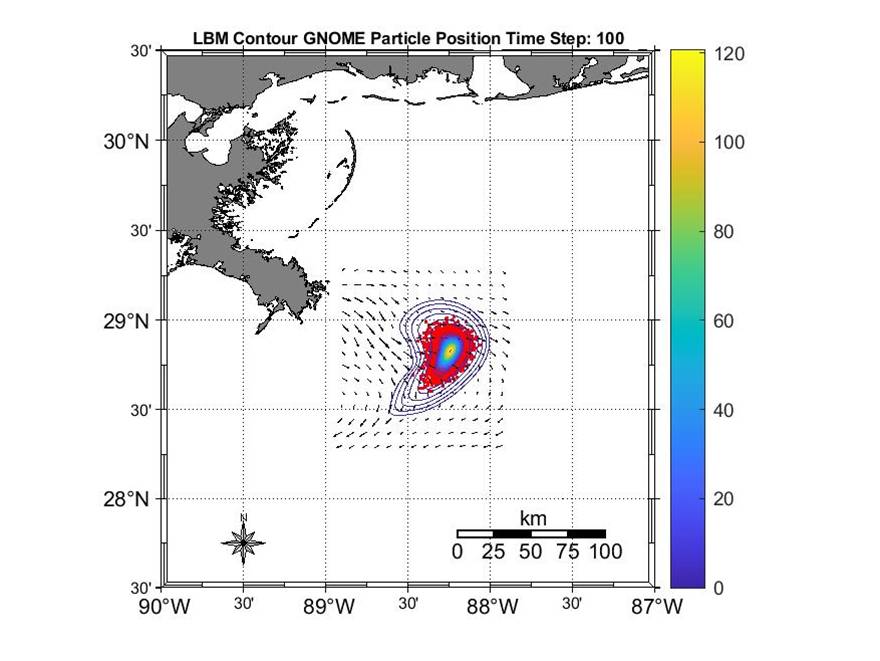} \hfill
    \includegraphics[width=.45\textwidth]{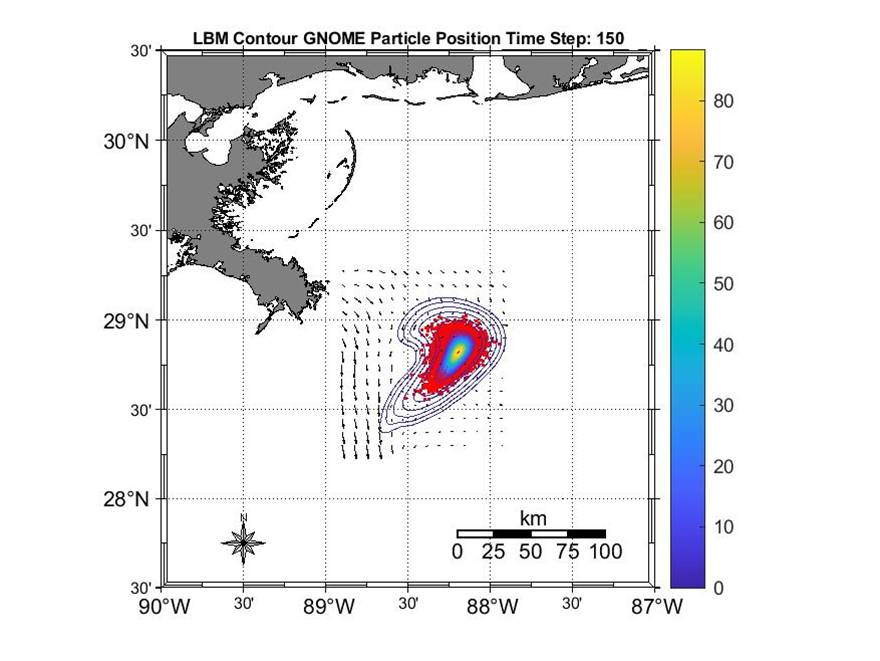} \hfill
    \caption{LBM oil density and GNOME particles with ocean current.}
    \label{fig:lbm-gnome-ocean}
}
\end{figure*}

We show the comparison results of  both LBM-ADE and GNOME in snap shots at time steps 2, 50, 100  and 150 in Figure \ref{fig:lbm-gnome-ocean} with background of ocean current in the study domain of Gulf of Mexico, where the contours show oil density in LBM and the red color particles are LEs in GNOME.   These times were chosen to omit the GNOME transient effect and the loss of concentration at the north eastern boundary. 

Figure \ref{fig:lbm-gnome-ocean} clearly shows that the contours can reach far out from the center mass of oil concentration where there are no presents of particles. Therefore, the LBM-ADE model is more effective to model the far-field impact of ocean oil spill to certain highly sensitive areas in terms of predicting how far a small amount of oil can reach and how long it takes. 

\subsection{Discussion of  GNOME Diffusion and LBM BGK Collision and Their Relation}

GNOME incorporates diffusion as horizontal eddy diffusivity in the water using the model of a random walk which theoretically corresponds to the solution of Advection-Dif\-fusion Equation (\ref{eq:lbm-ade-pde}) where $D$, referred to $D_g$, corresponds to  an isotropic diffusion \cite{noaa:technotes}:

\begin{quote}
"Diffusion is a way to capture [...] small scale circulations that [are] not captured in the underlying circulation model. This is often known as subgrid scale circulation. These small eddies [...]  serve to spread things out, or diffuse them." 
\end{quote}

To effectively represent diffusion during a spill the value is calibrated using over flight data \cite{noaa:gnome}. The process is somewhat of an art which, constrained by the current grid scale and geographic complexity,  strives to represent the spread without diluting the concentration.  The default value used by GNOME is $1 \times 10^5 \ cm^2/sec$.  

In our LBM simulation, diffusion  is incorporated into the model  through  the relaxation parameter, $\tau_g$, based on Equation (\ref{eq:lbm-ade-D}).   The physical scaling used in GNOME is used to calculate the equivalent $\tau_g$ from the value of $D$.  To avoid numerical instability, $\tau_g>0.50$.  While this limits the LBM for very small values of $D$, for modeling oil spills this does not present a significant limitation since for small $D$ the trajectory of the oil is driven primarily by the current .   In this analysis we have used various values given in the  Table~\ref{tab:diffusion-parameters}.  It shows the LBM-ADE model and its numerical solutions are robust enough to handle a wider range of values of $\tau_g$.

Table~\ref{tab:diffusion-parameters} shows the values of the corresponding diffusion for a non-dimensional Monte Carlo model,  GNOME, LBM-ADE and $\tau_g$.  While $D$ and $D_g$ are independent of the numerical stability of their respective models, $D_l$ is directly related to $\tau_g$ which contributes to the numerical stability of the LBM-ADE.    

\begin{table}[htb]
\centering
\caption{Diffusion Parameters in  Monte Carlo, GNOME and LBM-ADE Models.For this example the non-dimensional scaling parameters are:
length $(l=10.0)$, time $(t=2.0)$, and  velocity $(v=(0.0,-1.0)$.\label{tab:diffusion-parameters}}
\begin{tabular}{llll}
\hline
$D$ & $D_g(cm^2/s)$ & $D_l(l^2/lt)$ & $\tau_g(lt)$ \\ \hline

$0.28125$ & $2.61x10^6$ & $0.760134$ & $2.78040$ \\
$0.2112$5 & $1.96x10^6$ & $0.507945$ & $2.21284$ \\
$0.15125$ & $1.40x10^6$ & $0.408783$ & $1.72635$ \\
$0.10125$ & $9.04x10^5$ & $0.273648$ & $1.32095$ \\
$0.06125$ & $5.68x10^5$ & $0.165540$ & $0.99662$ \\
$0.03125$ & $2.90x10^5$ & $0.084459$ & $0.75338$ \\
$0.01125$ & $1.04x10^5$ & $0.030405$ & $0.59122$ \\
$0.00125$ & $1.16x10^4$ & $0.003378$ & $0.51014$ \\
\hline
\end{tabular}
\end{table}




    


\section{Complexity Analysis and Performance Evaluation}

In our study, both GNOME and LBM are ADE based models which provide numerical solutions that approximate the propagation of oil pollutants on ocean surface. In addition to evaluating their effectiveness of modeling ocean oil spill  far-field impacts, we conducted analytical analyses and empirical tests to compare their performance in terms of accuracy and computation complexity.  

\subsection{Accuracy of the LBM and GNOME} \label{lbm_gnome_accuracy}

Using an analysis based on the Chapman-Enskog expansion, Kruger \cite{kkkssv:lbm} shows that the Lattice Boltzmann Equation with the BGK collision operator (\ref{eq:lbm-ade-collison}) and quadratic equilibrium (\ref{eq:lbm-ade-geq})  recovers the ADE (\ref{eq:lbm-ade-pde}) up to $O(\Delta t ^2)$ and $O(\epsilon^2)$ where $\epsilon$ is the Knudsen number. Moreover, they show that the resulting error term is independent of the velocity.  In \cite{Zhang-Shi:2012}, T. Zhang, et al.  also analyze  the convergence of this LBM model and present numerical experiments which demonstrate the $O(\epsilon^2)$ convergence in space. Above reference \cite{kkkssv:lbm} also notes that the error term can be eliminated by introducing an artificial source term as shown in \cite{Chai-Zhao:2013}. In the same work \cite{Chai-Zhao:2013}, Chai and Zhao  show that  an LBM model with an artificial source term is exactly $O(\epsilon^2)$ convergent in space and the authors present numerical examples showing that their revised LBM model is slightly more accurate than \cite{Zhang-Shi:2012} for the same benchmark problem.

For our LBM model using a $(lx,ly)$ lattice in a $1^{\circ} \times 1^{\circ}$  computational domain the theoretical accuracy of the normalized density is $O(\epsilon^2)$, $1/(1^{\circ} \times 1^{\circ})$ which for $lx=ly=200$ is $2.5 \times 10^{-5}$.  In addition, smaller concentrations can be calculated in the far field limited by numerical round-off error, which may be smaller than the theoretical accuracy and provide valuable information.   


The "mover" components of GNOME which will be replaced by LBM are the LE trajectories and diffusion. As far as the trajectories,  GNOME uses a forward Euler method \cite{noaa:gnome} which is locally  $O(\Delta t ^2)$ \cite{Isaacson:1966, stoer:Numerical-Analysis} and globally $O(\Delta t)$.   As far as the diffusion,  GNOME uses classical diffusion, see Equation \ref{eq:lbm-ade-pde2},  which is modeled by a stochastic differential equation using a Monte Carlo method based on Brownian Motion.

\begin{equation}    \label{eq:lbm-ade-pde2}
\frac{\partial C}{\partial t}=D\nabla ^2 C
\end{equation}
 
The accuracy of the stochastic differential equation solved is $O(\sqrt\Delta t)$ \cite{Haugh:2017,Glasserman:2003,Gardiner:2009}, here the size of $\Delta t$ determines the degree of (non-)smoothness of the LE paths.  For evaluating {\em rare events} for comparison with LBM,  the density of the smallest detectable concentration is determined by the number of LEs rather than the smoothness of the path and time step. 

For our GNOME example, the spatial resolution of the concentration of particles is $(\frac 1 N$  $\frac A {\Delta x \Delta y})$ where $A$ is the study area and $\Delta x$ and $\Delta y$ are the grid size of the bins used for accumulating particles for the purpose of calculating densities.  We use $\Delta x=200,  \Delta y=200$ and an area of $1^{\circ} \times 1^{\circ}$ centered at (28.8,-88.4).  The smallest detectable non-zero probability of a particle in a bin is $1/N$ which is $1 \times 10^{-4}$. Throughout the far field where there are no particles the probability is constant equal to 0. 

\subsection{GNOME and LBM-ADE Computation Complexity Comparison}

GNOME  simulates  the  ocean oil  spill surface propagation  based  on  an ensemble of Lagrangian  particle trajectories which are combination of particle advection movements and stochastic Brownian motions with specified diffusion coefficient, $D_g$.  The basic computation unit is a Lagrangian element (LE) which consumes certain computation resources (CPU/GPU time and memory space) in each time step.  The total computation complexity in a given GNOME simulation is proportional to the total number of LEs and simulation time steps, $O(N_{le}) \times N_s$, where $N_{le}$ is the number of LEs and $N_s$ is the number of simulation steps.

On the other hand, LBM-ADE uses a kinetic process to simulate ocean oil spill surface propagation based on oil concentration density and movements on a lattice grid which are governed by a stream-collision equation with the BGK operator \eqref{eq:lbm-ade-collison}  and the equilibrium equation \eqref{eq:lbm-ade-geq}.  The LBM-ADE numerical solution is computed using an algorithm presented in Bao and Meskas \cite{bao:lbm-fluid}. In this case, the basic computation unit is a lattice node.    The total computation complexity in a given LBM-ADE simulation is proportional to the total number of lattice nodes and simulation time steps, $O(N_{l}^2) \times N_s$, where $N_{l}^2$ is the total number of lattice nodes in a 2D square lattice and $N_s$ is the number of simulation steps.

We conducted simulation tests to generate empirical data sets which are used to construct  computation complexity profiles for both GNOME and LBM-ADE models. For each simulation run, we use the same number of simulation steps (297 steps) in both GNOME and LBM-ADE models, so that we only need to consider the computation times related to total number LEs in GNOME and total number lattice nodes in LBM-ADE respectively.  We implemented a Python version of GNOME simulation program using pyGNOME API published by NOAA \cite{noaa:technotes} and we also implemented a Python version of the LBM-ADE simulation program.  Both simulation programs run on a Dell Latitude E6430 laptop computer with Intel dual-core CPU at 2.90 GHz, 8 GB RAM and 64-bit Windows 10 operating system. 

\begin{figure}[htb]
{
\centering
\includegraphics[width=0.45\textwidth]{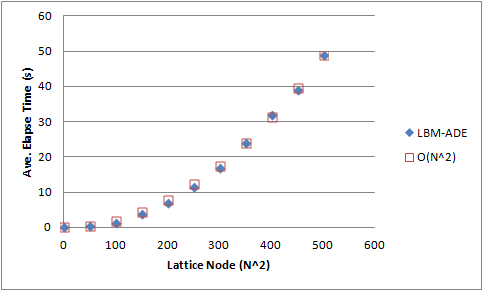}
\caption{LBM-ADE computation time profile.\label{fig:lbm_time}}
}
\end{figure}

Figure  \ref{fig:lbm_time} shows the LBM-ADE computation time profile as the lattice size ($N_l$) changes from 50 to 500 with incremental of 50.  The elapsed time is averaged over 10 simulation runs for each lattice size.  It shows the LBM-ADE computation time scaling agrees with the estimated scaling of $O(N_{l}^2)$.

\begin{figure}[htb]
{
\centering
\includegraphics[width=0.45\textwidth]{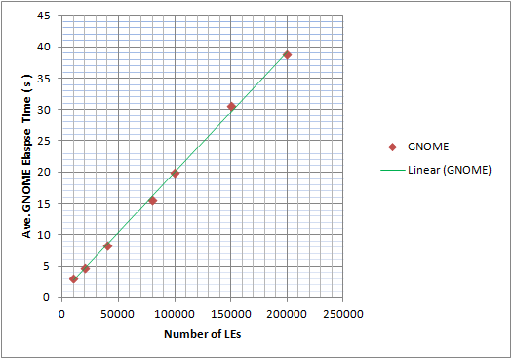}
\caption{GNOME computation time profile.\label{fig:gnome_time}}
}
\end{figure}

Figure  \ref{fig:gnome_time}  shows GNOME computation time profile as the number of LE changes from 10,000 to 200,000. The elapse time is averaged over 10 simulation runs for each total number of LEs  It shows the GNOME computation time is linear related to the total number of LEs.  

Referring to the example  mentioned in section \ref{lbm_gnome_accuracy}, the  LBM-ADE model with a (200 x 200) lattice  can achieve accuracy of the normalized density at level of $2.5 \times 10^{-5}$ and it takes 6.91 seconds as Figure  \ref{fig:lbm_time} shows.  For GNOME to reach the same level of accuracy, it would need 40000 LEs and it will take 8.40 seconds as Figure  \ref{fig:gnome_time}  shows. we found that, in this case,  the LBM-ADE  and the GNOME are comparable with LBM-ADE slightly outperforms GNOME in computation times. As Figure \ref{fig:far_field_benchmark} shows if we were estimating extremely low oil density at $1 \times 10^{-6}$ level, LBM-ADE will be far more advantage than GNOME in computation time.

\subsection{Potential Performance Gains in Parallel Processing}

Both LBM and GNOME algorithms lend themselves to parallel computing
techniques. Since each time step of the LBM algorithm can be
structured so that all calculations on nodes are locally dependent,
either depending only on the node or at most depending on nodes one
lattice edge apart, the algorithm is favorable to parallelization.

The explicit time step of the LBM process can be divided into two sub-steps (see sec. \ref{sec:lbm-ade-model}):

\begin{enumerate}
\item The first is comprised of the relaxation, the equilibrium and
  the macroscopic calculations which depend only on the node being
  updated.  Parallel domain partitioning techniques \cite{huelsemann:2006} enable simultaneous update of each
  node.

\item The second is streaming, which depends on the node being updated
  and the immediate neighboring nodes.  Parallel domain partitioning
  techniques and ghost cells for sub-domain boundaries enables
  parallel processing of each sub-domain simultaneously.
\end{enumerate}

Good speed-ups have been reported for multiple PUs (processing units)
by Körner C., et al \cite{koerner-etal:2006} for 64 CPU and linear
scale-up (fixed CPU load by scaling the domain size). In addition,
effective massive parallel techniques using GPGPUs (General Purpose
Graphics Processing Units) and CUDA are described by Jan{\ss}en and
Krafczyk \cite{janssen-krafczyk:2011} with "one order of magnitude
faster than comparable CPU implementations". Parallel threads
techniques using OpenMP and GPUs where used for an LBM solution of the Shallow Water Equations by Kevin Tubbs where linear speed-up was
obtained for up to 6 threads for OpenMPI \cite{tubbs:2010,tubbs-tsai:2010}.

Parallelizing each time step of the particle based algorithms in
GNOME is straightforward because each particle moves independently of each other particle.  These algorithms are often referred to as
{\em embarrassingly parallel}. Typical parallel techniques such as using
threads, multiple CPUs and message passing and GPU can be used with
various speed-up improvements \cite{anderson-etal:2013,lang-prehl:2017}.




\section{Conclusion}

We study and evaluate GNOME and the LBM-ADE model in their capability and effectiveness in terms of modeling ocean oil spill far-field impacts to certain sensitive areas. The results show that
GNOME is not necessarily the most appropriate tool to use  for analyzing far-field impact with very small oil densities.  
The LBM-ADE model, on the other hand, can provide numerical estimations to certain far-field area with very low oil density while GNOME cannot. In their performance evaluations, we find the LBM-ADE model performs better in accuracy and computation complexity. For these reasons, the LBM-ADE model is a viable alternative to GNOME in modeling the far-field impacts to highly sensitive areas. 

For future research, We will explore the possibilities of using LBM-NSE as well as LBM-ADE to model multi species and multi phase flows. This will enhance our ability to model the ocean oil weathering process, such as mixed water and oil droplets in ocean water columns and their far-field impacts to marine lives beneath ocean surface.

\section{Appendix: Estimating Limits of Monte-Carlo Simulations}

In some special cases, it is possible to evaluate the limits of Monte-Carlo simulations exactly. The results of this analysis can be used to create benchmark situtations for the numerical algorithms. To illustrate one particular approach, we consider the simple case of a constant advection field in $y$-direction with $v_y<0$ such that the oil particles will be advected downward and no advection in the $x$-direction ($v_x=0$). In this case, we can solve the advection-diffusion equation explicitly. For a $\delta$-initial condition corresponding to a release of the oil at the origin, the probability density of finding a particle at a location $(x,y)$ after a time $T$ is given by
\begin{equation}
    p(x,y,T) = \frac{1}{2\pi\sigma^2T}{\mathrm{e}}^{-(x^2+(y-v_yT)^2)/(2\sigma^2T)}
\end{equation}
From there, we can compute the probability of finding a particle above $y>L>0$. For typical parameter values, corresponds to a rare event as the advection is pulling the particle downwards and only a few particles will be able to reach a point with $y>L$. The exact probability $P$ is found from the probability density above as
\begin{eqnarray}
    P(T) &=& \int_{-\infty}^{\infty} dx\int_L^{\infty}p(x,y,T),\,dy  \nonumber \\ 
         &=& 1-\Phi\left(\frac{L-v_yT}{\sigma\sqrt{T}}\right)
\end{eqnarray}
Here, $\Phi$ denotes the cumulative normal distribution. This result can be interpreted as follows: For very small times, it is unlikely for particles to reach the region $y>L>0$ simply because the particles need time to diffuse away from the source.
\begin{figure}[htb]
{
\centering
\includegraphics[width=0.35\textwidth]{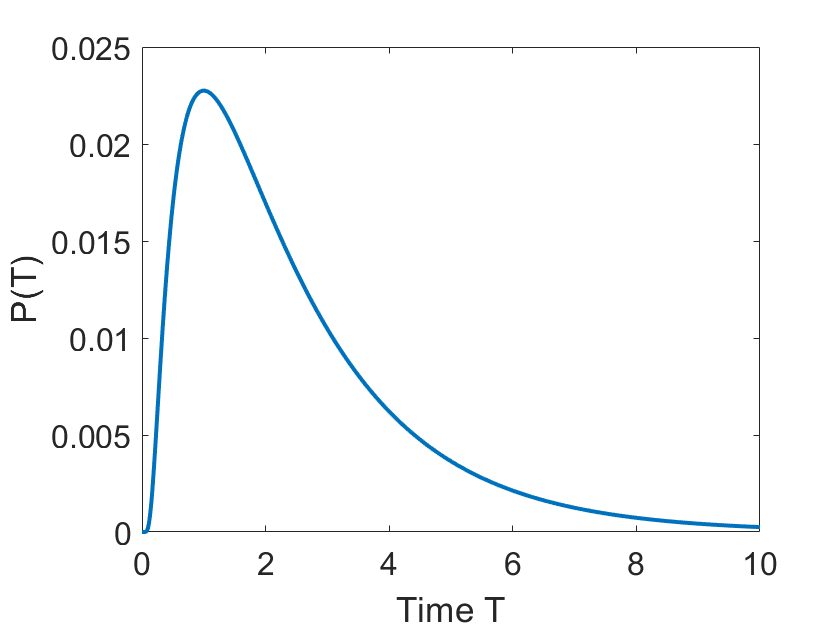}
\caption{Example illustrating the probability $P(T)$. Here, $v_y = -1$, $\sigma = 1$, and $L = 1$. The time $T$ ranges from 0 to 10 .\label{fig:prob_T}}
}
\end{figure}
As the time increases, particles leave the source and they are being diffused and advected. Most particles are moved downward by the drift $v_y<0$, away from the region $y>L>0$. However, for some particles corresponding to a few rare events, the diffusion happens to act against the drift and there is a chance for those particles to reach this region. As time grows, however, it becomes again less and less likely for the particles to reach this region since the drift is constantly dragging them downward. Figure \ref{fig:prob_T} illustrates this dependence of the probability on the time $T$.

Assume now that we run a Monte-Carlo simulation with $N_{max}$ particles. Clearly, if we expect less than one particle to reach the area with $y>L$, we will not be able to rely on Monte-Carlo simulations to collect sufficient statistics about that region. Therefore, setting $P= 1/N_{max}$ gives us an estimate for the limit of the Monte-Carlo method. Using the formula above, we can compute the maximum distance $L$. Solving for $L$ we obtain
\begin{equation}
    L = v_yT + \sigma\sqrt{T}\Phi^{-1}\left(1-\frac{1}{N_{max}}\right)
\end{equation}
where $\Phi^{-1}$ denotes the inverse cumulative normal distribution. Note that $L=L(T)$ is still a function of the time $T$. However, it is easy to see that, as a function of $T$, its maximum is given by
\begin{equation}
    L_{max} = \frac{1}{4|v_y|}\sigma^2\left(\Phi^{-1}\left(1-\frac{1}{N_{max}}\right)\right)^2
\end{equation}
which allows us to estimate the limit of Monte-Carlo simulations in this particular case.

\section*{Acknowledgments}
Computational support was provided by The City University of New York High Performance Computing Center, which is operated by the College of Staten Island funded, in part, by the National Science Foundation grants CNS-0958379 and CNS-0855217, and DMS-2012548. This pro\-ject was also supported, in part, by the PSC-CUNY Research Award 63570-00 51. 
We thank Robyn N. Conmy, PhD. US EPA Office of Research and Development, Remediation Pollution Control Division for her invaluable comments and guidance in understanding environmental transport and weathering of oil.  We further thank Andrew Poje, PhD. Professor of Mathematics, College of Staten Island for his assistance with UWIN-CM.

\bibliographystyle{cas-model2-names}

\bibliography{cas-refs}

\begin{thebibliography}{38}
\expandafter\ifx\csname natexlab\endcsname\relax\def\natexlab#1{#1}\fi
\providecommand{\url}[1]{\texttt{#1}}
\providecommand{\href}[2]{#2}
\providecommand{\path}[1]{#1}
\providecommand{\DOIprefix}{doi:}
\providecommand{\ArXivprefix}{arXiv:}
\providecommand{\URLprefix}{URL: }
\providecommand{\Pubmedprefix}{pmid:}
\providecommand{\doi}[1]{\href{http://dx.doi.org/#1}{\path{#1}}}
\providecommand{\Pubmed}[1]{\href{pmid:#1}{\path{#1}}}
\providecommand{\bibinfo}[2]{#2}
\ifx\xfnm\relax \def\xfnm[#1]{\unskip,\space#1}\fi
\bibitem[{Anderson et~al.(2013)Anderson, Jankowski, Grubb, Engel and
  Glotzer}]{anderson-etal:2013}
\bibinfo{author}{Anderson, J.A.}, \bibinfo{author}{Jankowski, E.},
  \bibinfo{author}{Grubb, T.L.}, \bibinfo{author}{Engel, M.},
  \bibinfo{author}{Glotzer, S.C.}, \bibinfo{year}{2013}.
\newblock \bibinfo{title}{Massively parallel {Monte} {Carlo} for many-particle
  simulations on {GPUs}}.
\newblock \bibinfo{journal}{Journal of Computational Physics}
  \bibinfo{volume}{254}, \bibinfo{pages}{27 -- 38}.
\bibitem[{Banda and Seaid(2012)}]{banda:lbm-shallow-water}
\bibinfo{author}{Banda, M.K.}, \bibinfo{author}{Seaid, M.},
  \bibinfo{year}{2012}.
\newblock \bibinfo{title}{Lattice {Boltzmann} simulation for shallow water flow
  applications}, in: \bibinfo{editor}{Zheng, J.} (Ed.),
  \bibinfo{booktitle}{Hydrodynamics}. \bibinfo{publisher}{IntechOpen},
  \bibinfo{address}{Rijeka}. chapter~\bibinfo{chapter}{11}.
\bibitem[{Bao and Meskas(2011)}]{bao:lbm-fluid}
\bibinfo{author}{Bao, Y.B.}, \bibinfo{author}{Meskas, J.},
  \bibinfo{year}{2011}.
\newblock \bibinfo{title}{Lattice {Boltzmann} Method for Fluid Simulations}.
\newblock \bibinfo{type}{Technical Report}. Courant Institute of Mathematical
  Science, New York University. \bibinfo{address}{New York}.
\bibitem[{Bodkin et~al.(2014)Bodkin, Esler, Rice, Matkin and
  Ballachey}]{bodkin_esler_rice_matkin_ballachey:2014}
\bibinfo{author}{Bodkin, J.L.}, \bibinfo{author}{Esler, D.},
  \bibinfo{author}{Rice, S.D.}, \bibinfo{author}{Matkin, C.O.},
  \bibinfo{author}{Ballachey, B.E.}, \bibinfo{year}{2014}.
\newblock \bibinfo{title}{The effects of spilled oil on coastal ecosystems:
  lessons from the Exxon Valdez spill}. \bibinfo{publisher}{Cambridge
  University Press}.
\newblock Conservation Biology, pp. \bibinfo{pages}{311--346}.
\bibitem[{Cercignani(1988)}]{cercignani:bgk}
\bibinfo{author}{Cercignani, C.}, \bibinfo{year}{1988}.
\newblock \bibinfo{title}{The {Boltzmann} Equation and Its Applications}.
\newblock \bibinfo{publisher}{Springer}, \bibinfo{address}{New York}.
\bibitem[{Chai and Zhao(2013)}]{Chai-Zhao:2013}
\bibinfo{author}{Chai, Z.}, \bibinfo{author}{Zhao, T.S.}, \bibinfo{year}{2013}.
\newblock \bibinfo{title}{Lattice {Boltzmann} model for the
  convection-diffusion equation}.
\newblock \bibinfo{journal}{Physical Review E} \bibinfo{volume}{87}.
\bibitem[{Csanady(1973)}]{csanady:diffusion}
\bibinfo{author}{Csanady, G.T.}, \bibinfo{year}{1973}.
\newblock \bibinfo{title}{Turbulent Diffusion in the environment}.
\newblock \bibinfo{publisher}{Dordrecht, Holland: D. Reidel Publishing
  Company}, \bibinfo{address}{Netherlands}.
\bibitem[{Dedits et~al.(2015)Dedits, Poje, T.Schäfer and
  Vukadinovic}]{dpsv:lbm-ade}
\bibinfo{author}{Dedits, E.}, \bibinfo{author}{Poje, A.},
  \bibinfo{author}{T.Schäfer}, \bibinfo{author}{Vukadinovic, J.},
  \bibinfo{year}{2015}.
\newblock \bibinfo{title}{Averaging and spectral properties for the 2d
  advection-diffusion equation in the semi-classical limit for vanishing
  diffusivity}.
\newblock \bibinfo{journal}{Physics D} \bibinfo{volume}{310},
  \bibinfo{pages}{1--18}.
\bibitem[{Dipper and Chua(2004)}]{IPIECA:2004}
\bibinfo{author}{Dipper, F.}, \bibinfo{author}{Chua, T.E.},
  \bibinfo{year}{2004}.
\newblock \bibinfo{title}{Biological Impacts of Oil Pollution: Fisheries}.
\newblock \bibinfo{type}{Technical Report}. International Petroleum Industry
  Environmental Conservation Association (IPIECA).
  \bibinfo{address}{London,United Kingdom}.
\bibitem[{Gardiner(2009)}]{Gardiner:2009}
\bibinfo{author}{Gardiner, C.}, \bibinfo{year}{2009}.
\newblock \bibinfo{title}{Stochastic Methods A Handbook for the Natural and
  Social Sciences}.
\newblock \bibinfo{publisher}{Springer-Verlag}, \bibinfo{address}{Berlin
  Heidelberg}.
\bibitem[{Glasserman(2003)}]{Glasserman:2003}
\bibinfo{author}{Glasserman, P.}, \bibinfo{year}{2003}.
\newblock \bibinfo{title}{Monte Carlo Methods in Financial Engineering}.
\newblock \bibinfo{publisher}{Springer}, \bibinfo{address}{New York}.
\bibitem[{Gracia et~al.(2020)Gracia, Murawski and Vázquez-Bader}]{Gracia:2020}
\bibinfo{author}{Gracia, A.}, \bibinfo{author}{Murawski, S.},
  \bibinfo{author}{Vázquez-Bader, A.}, \bibinfo{year}{2020}.
\newblock \bibinfo{title}{Impacts of Deep Oil Spills on Fish and Fisheries}.
\newblock pp. \bibinfo{pages}{414--430}.
\bibitem[{Graham and Reilly(2011)}]{graham:bp2011}
\bibinfo{author}{Graham, R.}, \bibinfo{author}{Reilly, W.K.},
  \bibinfo{year}{2011}.
\newblock \bibinfo{title}{Deep Water: The Gulf Oil Disaster And The Future Of
  Offshore Drilling - Report to the President}.
\newblock \bibinfo{type}{Technical Report}. National Commission on the BP Deep
  Water Horizon Oil Spill and Offshore Drilling. \bibinfo{address}{Washington,
  DC}.
\bibitem[{Ha and Ku(2012)}]{haku:lbmoil}
\bibinfo{author}{Ha, S.}, \bibinfo{author}{Ku, N.}, \bibinfo{year}{2012}.
\newblock \bibinfo{title}{Lattice {Boltzmann} simulation for the prediction of
  oil slick movement and spread in ocean environment}, in:
  \bibinfo{booktitle}{Proceeding of 2012 International Offshore and Polar
  Engineering Conference}, \bibinfo{address}{Rhodes, Greece}. pp.
  \bibinfo{pages}{783--788}.
\bibitem[{H{\"u}lsemann et~al.(2006)H{\"u}lsemann, Kowarschik, Mohr and
  R{\"u}de}]{huelsemann:2006}
\bibinfo{author}{H{\"u}lsemann, F.}, \bibinfo{author}{Kowarschik, M.},
  \bibinfo{author}{Mohr, M.}, \bibinfo{author}{R{\"u}de, U.},
  \bibinfo{year}{2006}.
\newblock \bibinfo{title}{Parallel geometric multigrid}, in:
  \bibinfo{editor}{Bruaset, A.M.}, \bibinfo{editor}{Tveito, A.} (Eds.),
  \bibinfo{booktitle}{Numerical Solution of Partial Differential Equations on
  Parallel Computers}, \bibinfo{publisher}{Springer Berlin Heidelberg},
  \bibinfo{address}{Berlin, Heidelberg}. pp. \bibinfo{pages}{165--208}.
\bibitem[{Isaacson and Keller(1966)}]{Isaacson:1966}
\bibinfo{author}{Isaacson, E.}, \bibinfo{author}{Keller, H.B.},
  \bibinfo{year}{1966}.
\newblock \bibinfo{title}{Analysis of Numerical Methods}.
\newblock \bibinfo{publisher}{Wiley}, \bibinfo{address}{New York}.
\bibitem[{Janßen and Krafczyk(2011)}]{janssen-krafczyk:2011}
\bibinfo{author}{Janßen, C.}, \bibinfo{author}{Krafczyk, M.},
  \bibinfo{year}{2011}.
\newblock \bibinfo{title}{Free surface flow simulations on {GPGPUs} using the
  {LBM}}.
\newblock \bibinfo{journal}{Computers \& Mathematics with Applications}
  \bibinfo{volume}{61}, \bibinfo{pages}{3549 -- 3563}.
\newblock \bibinfo{note}{{Mesoscopic} Methods for Engineering and Science —
  Proceedings of ICMMES-09}.
\bibitem[{K{\"o}rner et~al.(2006)K{\"o}rner, Pohl, R{\"u}de, Th{\"u}rey and
  Zeiser}]{koerner-etal:2006}
\bibinfo{author}{K{\"o}rner, C.}, \bibinfo{author}{Pohl, T.},
  \bibinfo{author}{R{\"u}de, U.}, \bibinfo{author}{Th{\"u}rey, N.},
  \bibinfo{author}{Zeiser, T.}, \bibinfo{year}{2006}.
\newblock \bibinfo{title}{{Parallel Lattice Boltzmann Methods} for {CFD
  Applications}}, in: \bibinfo{editor}{Bruaset, A.M.}, \bibinfo{editor}{Tveito,
  A.} (Eds.), \bibinfo{booktitle}{Numerical Solution of Partial Differential
  Equations on Parallel Computers}, \bibinfo{publisher}{Springer Berlin
  Heidelberg}, \bibinfo{address}{Berlin, Heidelberg}. pp.
  \bibinfo{pages}{439--466}.
\bibitem[{Kruger et~al.(2017)Kruger, Kusumaatmaja, Kuzmin, Shardt, Silva and
  Viggen}]{kkkssv:lbm}
\bibinfo{author}{Kruger, T.}, \bibinfo{author}{Kusumaatmaja, H.},
  \bibinfo{author}{Kuzmin, A.}, \bibinfo{author}{Shardt, O.},
  \bibinfo{author}{Silva, G.}, \bibinfo{author}{Viggen, E.M.},
  \bibinfo{year}{2017}.
\newblock \bibinfo{title}{{T}he {L}attice {B}oltzmann {M}ethod - {P}rinciples
  and {P}ractice}.
\newblock \bibinfo{publisher}{Springer}, \bibinfo{address}{Cham, Switzerland}.
\bibitem[{Lang and Prehl(2017)}]{lang-prehl:2017}
\bibinfo{author}{Lang, J.}, \bibinfo{author}{Prehl, J.}, \bibinfo{year}{2017}.
\newblock \bibinfo{title}{An embarrassingly parallel algorithm for random walk
  simulations on random fractal structures}.
\newblock \bibinfo{journal}{Journal of Computational Science}
  \bibinfo{volume}{19}, \bibinfo{pages}{1 -- 10}.
\bibitem[{Li et~al.(2017)Li, Mei and Klausner}]{lmk:lbmcde}
\bibinfo{author}{Li, L.}, \bibinfo{author}{Mei, R.}, \bibinfo{author}{Klausner,
  J.F.}, \bibinfo{year}{2017}.
\newblock \bibinfo{title}{Lattice {Boltzmann} models for the
  convection-diffusion equation: {D2Q5} vs {D2Q9}}.
\newblock \bibinfo{journal}{International Journal of Heat and Mass Transfer}
  \bibinfo{volume}{108}, \bibinfo{pages}{41--62}.
\bibitem[{Li and Huang(2009)}]{lihuang:lbm-shallow-water}
\bibinfo{author}{Li, Y.}, \bibinfo{author}{Huang, P.}, \bibinfo{year}{2009}.
\newblock \bibinfo{title}{A coupled lattice {Boltzmann} model for the shallow
  water-contamination system}.
\newblock \bibinfo{journal}{International Journal for Numerical Methods in
  Fluids} \bibinfo{volume}{59}, \bibinfo{pages}{195--213}.
\bibitem[{{Martin Haugh}(2017)}]{Haugh:2017}
\bibinfo{author}{{Martin Haugh}}, \bibinfo{year}{2017}.
\newblock \bibinfo{title}{Monte-Carlo Simulation}.
\newblock \bibinfo{type}{Technical Report}. Columbia University.
  \bibinfo{address}{New York, New York}.
\bibitem[{Maslo et~al.(2014)Maslo, Penjan and Zagar}]{maslo:lbmoill}
\bibinfo{author}{Maslo, A.}, \bibinfo{author}{Penjan, J.},
  \bibinfo{author}{Zagar, D.}, \bibinfo{year}{2014}.
\newblock \bibinfo{title}{Large-scale oil spill simulation using the lattice
  {Boltzmann} method, validation on the {Lebanon} oil spill case}.
\newblock \bibinfo{journal}{Marine Pollution Bulletin} \bibinfo{volume}{84},
  \bibinfo{pages}{225--235}.
\bibitem[{Mishra and Kumar(2015)}]{mishra:oilspillwethering}
\bibinfo{author}{Mishra, A.K.}, \bibinfo{author}{Kumar, G.S.},
  \bibinfo{year}{2015}.
\newblock \bibinfo{title}{Weathering of oil spill: Modelling and analysis}, in:
  \bibinfo{editor}{Dwarakish, G.} (Ed.), \bibinfo{booktitle}{Proceedings of the
  International Conference on Water Resources, Coastal and Ocean Engineering}.
  \bibinfo{publisher}{Elsevier Procedia}, pp. \bibinfo{pages}{435--442}.
\bibitem[{Munk(1950)}]{munk:oceanmodell}
\bibinfo{author}{Munk, W.H.}, \bibinfo{year}{1950}.
\newblock \bibinfo{title}{On the wind-driven ocean circulation}.
\newblock \bibinfo{journal}{Journal of Meteorology} \bibinfo{volume}{7},
  \bibinfo{pages}{79--93}.
\bibitem[{{NOAA OR\&R}(2012)}]{noaa:gnome}
\bibinfo{author}{{NOAA OR\&R}}, \bibinfo{year}{2012}.
\newblock \bibinfo{title}{General NOAA Operational Modeling Environment (GNOME)
  Technical Documentation}.
\newblock \bibinfo{type}{Technical Report}. NOAA, Office of Response and
  Restoration. \bibinfo{address}{Seattle, Washington}.
\bibitem[{{NOAA OR\&R}(2021)}]{noaa:technotes}
\bibinfo{author}{{NOAA OR\&R}}, \bibinfo{year}{2021}.
\newblock \bibinfo{title}{{GNOME} tech notes}.
\newblock
  \bibinfo{howpublished}{\url{https://gnome.orr.noaa.gov/doc/pygnome/tech_notes.html?highlight=diffusion}}.
\newblock \bibinfo{note}{Accessed: 2021-03-22}.
\bibitem[{Nuraiman(2017)}]{nuraiman:lbm-ocean}
\bibinfo{author}{Nuraiman, D.}, \bibinfo{year}{2017}.
\newblock \bibinfo{title}{Modeling and simulation of ocean wave propagation
  using lattice {Boltzmann} method}.
\newblock \bibinfo{journal}{Journal of Physics} \bibinfo{volume}{Conf. Ser.
  893}, \bibinfo{pages}{25--29}.
\bibitem[{Stoer and Bulirsch(2002)}]{stoer:Numerical-Analysis}
\bibinfo{author}{Stoer, J.}, \bibinfo{author}{Bulirsch, R.},
  \bibinfo{year}{2002}.
\newblock \bibinfo{title}{Introduction to Numerical Analysis}.
\newblock \bibinfo{publisher}{Springer}, \bibinfo{address}{New York}.
\bibitem[{Sukop and Thorne(2007)}]{sukop:lbm-geoscientist}
\bibinfo{author}{Sukop, M.C.}, \bibinfo{author}{Thorne, D.T.},
  \bibinfo{year}{2007}.
\newblock \bibinfo{title}{Lattice Boltzmann Modeling – An Introduction for
  Geoscientists and Engineers}.
\newblock \bibinfo{publisher}{Springer}, \bibinfo{address}{Switzerland}.
\bibitem[{Tubbs(2010)}]{tubbs:2010}
\bibinfo{author}{Tubbs, K.}, \bibinfo{year}{2010}.
\newblock \bibinfo{title}{Lattice {Boltzmann} modeling for shallow water
  equations using high performance computing}.
\newblock \bibinfo{publisher}{Louisiana State University Digital Commons}.
\bibitem[{Tubbs and Tsai(2010)}]{tubbs-tsai:2010}
\bibinfo{author}{Tubbs, K.R.}, \bibinfo{author}{Tsai, F.T.C.},
  \bibinfo{year}{2010}.
\newblock \bibinfo{title}{{GPU} accelerated lattice {Boltzmann} model for
  shallow water flow and mass transport}.
\newblock \bibinfo{journal}{International Journal for Numerical Methods in
  Engineering} \bibinfo{volume}{86}, \bibinfo{pages}{316--334}.
\bibitem[{{University of Washington}((last access April 12th,
  2020))}]{uwin-cm:ocean-model}
\bibinfo{author}{{University of Washington}}, \bibinfo{year}{(last access April
  12th, 2020)}.
\newblock \bibinfo{title}{The unified wave interface-coupled model}.
\newblock \bibinfo{note}{\url{https://orca.atmos.washington.edu/models.php}}.
\bibitem[{Wolf-Gladrow(2005)}]{gladrow:lbm}
\bibinfo{author}{Wolf-Gladrow, D.A.}, \bibinfo{year}{2005}.
\newblock \bibinfo{title}{{L}attice-{G}as {C}ellular {A}utomata and {L}attice
  {B}oltzmann {M}odels - An Introduction}.
\newblock \bibinfo{publisher}{Springer}, \bibinfo{address}{Berlin}.
\bibitem[{Zhang et~al.(2012)Zhang, Shi, Guo, Chai and Lu}]{Zhang-Shi:2012}
\bibinfo{author}{Zhang, T.}, \bibinfo{author}{Shi, B.}, \bibinfo{author}{Guo,
  Z.}, \bibinfo{author}{Chai, Z.}, \bibinfo{author}{Lu, J.},
  \bibinfo{year}{2012}.
\newblock \bibinfo{title}{General bounce-back scheme for concentration boundary
  condition in the lattice-{Boltzmann} method}.
\newblock \bibinfo{journal}{Physical Review E} \bibinfo{volume}{85}.
\bibitem[{Zhang et~al.(2020a)Zhang, Schaefer and Kress}]{zzhang:wsc2020}
\bibinfo{author}{Zhang, Z.}, \bibinfo{author}{Schaefer, T.},
  \bibinfo{author}{Kress, M.E.}, \bibinfo{year}{2020}a.
\newblock \bibinfo{title}{A lattice {Boltzmann} advection diffusion model for
  ocean oil spill surface transport prediction}, in:
  \bibinfo{booktitle}{Proceedings of the Winter Simulation Conference},
  \bibinfo{organization}{Winter Simulation Conference, December 14th-16th,
  Virtual Conference}.
\bibitem[{Zhang et~al.(2020b)Zhang, Schaefer and Kress}]{zzhang:ams2020}
\bibinfo{author}{Zhang, Z.}, \bibinfo{author}{Schaefer, T.},
  \bibinfo{author}{Kress, M.E.}, \bibinfo{year}{2020}b.
\newblock \bibinfo{title}{{Lattice Boltzmann Method for Ocean Oil Spill
  Propagation Model and Simulation - A Comparison Study of Navier Stokes Model
  and Advection Diffusion Model}}, in: \bibinfo{booktitle}{Proceedings of the
  100th American Meteorological Society Annual Meeting},
  \bibinfo{organization}{100th American Meteorological Society Annual Meeting,
  January 12th-16th, Boston, MA}.

\end{thebibliography}


\bio{figs/zzhang}
 ZHANYANG ZHANG is an associate professor in the Ph. D. Program in Computer Science  at  The  Graduate  Center  of  the City  University  of  New  York  and  in  the  Computer  Science  Department  at  the  College  of  Staten  Island.   He earned his Ph.D. in Computer Science from City University of New York. His research interests include wireless networks, numerical modeling and simulations, IoT/sensor networks in smart city applications, high performance and cloud computing. His email address is \email{zhanyang.zhang@csi.cuny.edu}.
His website is
\href{http://www.cs.csi.cuny.edu/~zhangz/}{http://www.cs.csi.cuny.edu/$\sim$zhangz/}.
\endbio

\bio{figs/tschaefer}
TOBIAS SCH{\"A}FER  is a professor in the Ph. D. Program in Physics at The Graduate Center of the City University of New York and in the Department of Mathematics at the College of Staten Island. He earned his Ph.D. in Theoretical Physics at the University of D{\"u}sseldorf, Germany. His research interests include fluid dynamics, turbulence, and nonlinear optics. His email address is \email{tobias@math.csi.cuny.edu}. His website is 
\href{http://www.math.csi.cuny.edu/~tobias}{http://www.math.csi.cuny.edu/$\sim$tobias}. 
\endbio

\vspace{3in}

\bio{figs/MKress}
MICHAEL KRESS is an Emeritus Professor in the Ph. D. Program in Computer Science at The Graduate Center of the City University of New York and in the Computer Science Department at the College of Staten Island.   He earned his Ph.D. in Magneto-hydrodynamics from The Courant Institute of Mathematical Sciences, of New York University.  His research interests include high performance computing and numerical modeling and simulation of interdisciplinary systems including, environmental science applications regarding hurricanes, pollutant transport and transportation modeling as well as psychology and behavioral science using social network analysis.  He founded the City University of New York High Performance Computing Center.  His email address is \email{Michael.Kress@csi.cuny.edu}. 
\endbio

\end{document}